# Gemini MCAO Control System

C. Boyer, J. Sebag, B. Ellerbroek, Gemini Observatory, Hilo, HI 96720, USA

Abstract

The Gemini Observatory is planning to implement a Multi Conjugate Adaptive Optics (MCAO) System as a facility instrument for the Gemini-South telescope. The system will include 5 Laser Guide Stars, 3 Natural Guide Stars, and 3 Deformable mirrors optically conjugated at different altitudes to achieve near-uniform atmospheric compensation over a 1 arc minute square field of view [1,2,3,4]. The control of such a system will be split into 3 main functions: the control of the opto-mechanical assemblies of the whole system (including the Laser, the Beam Transfer Optics and the Adaptive Optics bench), the control of the Adaptive Optics System itself at a rate of 800FPS and the control of the safety system. The control of the Adaptive Optics System is the most critical in terms of real time performances. The control system will be an EPICS based system. In this paper, we will describe the requirements for the whole MCAO control system, preliminary designs for the control of the opto-mechanical devices and architecture options for the control of the Adaptive Optics system and the safety system.

## 1 OVERVIEW

As for conventional Laser Guide Star (LGS) Adaptive Optics (AO) system, the MCAO system will contain the six primary subsystems as follows: (1) the Laser System (LS), which includes all the elements necessary to produce the 5 laser guide stars; (2) the Beam Transfer Optics (BTO) used to transmit the laser light to the Laser Launch telescope; (3) the Laser Launch Telescope (LLT) located behind the secondary mirror at the top end of the Gemini Telescope; (4) the Adaptive Optics Module (AOM), which includes the deformable mirrors, the wave front sensors and the associated optical, mechanical, electrical components; (5) the Safe Aircraft Localization and Satellite Avoidance System (SALSA); (6) the Control System (CS), which will implement the Real Time Controller (RTC), that drives the deformable mirror based upon wavefront sensor measurements. This system also provides the supporting control functions such as opening/closing control loops, and will implement the control of all the opto-mechanical devices and loops of the BTO, LLT, AOM as well as the control of the SALSA system.

The MCAO CS controls the alignment, operation, and diagnostics of the whole MCAO system. It must manage a large number of opto-mechanical devices and still meet stringent real-time performance requirements. In order to do so the MCAO CS is split in 3 main functions: (i) the control of the Adaptive Optics System, this control is achieved by the RTC, which performs the real time wave front reconstruction and directly controls the deformable mirrors (DM) tip-tilt mirror (TTM), and the readout of the WFS components; (ii) the control of the SALSA system, (iii), the control of the all the opto-mechanical assemblies, which is implemented in 3 controllers:

- The Adaptive Optics Module Controller, which manages all of the opto-mechanical assemblies of the AOM except the deformable mirrors, the tip-tilt mirror and the readout of the WFS.
- The Beam Transfer / Laser Launch Telescope Controller which manages all of the opto-mechanical assemblies of the BTO and LLT.
- The Laser Controller, which manages the opto-mechanical assemblies of the laser itself (not described in this paper).

On top of this a sequencer component will be implemented to manage all the independent subsystems and act as the main public interface for the entire MCAO system. The sequencer will coordinate all of the internal tasks and provide external systems with the commands and status information they need to control the MCAO System.

The MCAO CS will be implemented using the standard Gemini Control System model. It will be a sub-system of the Observatory Control System and fully implemented as an EPICS system.

## 2 CONTROL OF THE AO SYSTEM

### 2.1 Requirements

This controller is dedicated to the Adaptive Optics control loop itself. It is the heart of the system and the most critical part in terms of real time performance. It will handle 3 basic real time functions: (i) the Natural Guide Star (NGS) real time control process; which controls the 3 tip/tilt sensors, the TTM and 3 DM anisoplanatism modes at a rate of 800Hz (Number of operations required: 3.16Mflops) (ii) the Laser Guide

Star (LGS) real time control process; which controls the 5 16x16 Shack Hartmann Wavefront Sensors (total of 2040 illuminated sub-apertures) and the 3 DMs (a total of 636 active actuators and 422 unactive actuators) at a rate of 800Hz (Number of operations required 2.26Gflops) (iii) the optimization and background processes, the goal of such processes is to continuously optimize or update the different parameters of the closed loop processes according to the atmospheric conditions for example, and also to provide data to outside components such as the MCAO BTO system.

### 2.2 Hardware solution

As discussed in the previous paragraph, the LGS control is the major user of CPU power, with a matrix multiplication being the most critical part. Performance at a level of ~3Gflops is required. Results of studies on current processors and architectures as well as benchmarks have both converged to a G4 PPC solution.

Such a solution is presented here. It is based on 4 VME Synergy quad PPC VSS4 boards (1 Synergy board per DM). The VSS4 board has one PMC (PCI Mezzanine Card) site, and using Synergy Micro Systems PEX3 PMC expander takes that site to yield three more, for a total of 3 PMC sites per VSS4 motherboard. To output the signals to the DM, a high speed parallel interface (PIO) board plugged directly into one of the PMC site of each quad G4 board will be used. To input the pixels, we will use a second high speed parallel interface board plugged into another PMC site. Each Synergy board will receive the pixels values, compute the centroids, perform the matrix multiplication that corresponds to its DM and send the actuator voltage through its daughter board directly. The last VSS4 board is dedicated to the optimization process, the control of the NGS process and the EPICS interface.

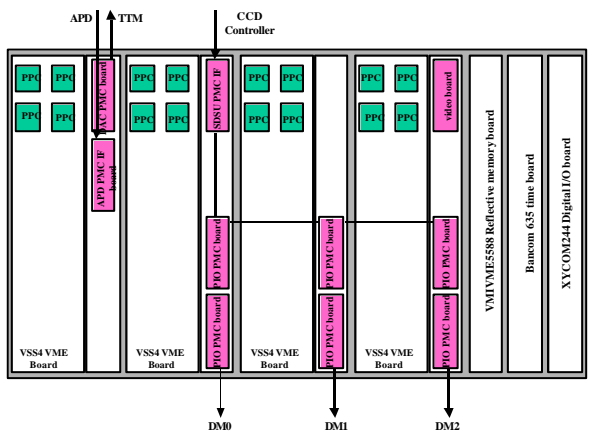

Figure1: RTC possible hardware architecture

The RTC will be developed outside Gemini. The RTC Vendor will define the final design of the RTC architecture.

## 3. CONTROL OF THE SALSA SYSTEM

The SALSA system consists of 3 main functions: (i) the Laser Traffic Control System (LTCS), for the beam collision avoidance; (ii) the aircraft detection; (iii) the satellite avoidance.

The LTCS will collect pointing information from all the telescopes located on the site, will determine in real time if collisions will occur and will close the safety shutter of the Laser Telescope in case of collision. The LTCS is developed jointly by the Keck Observatory and Gemini Observatory.

There will be 3 levels of aircraft detection:

- One infrared camera mounted on the telescope with a small field of view. This camera is boresited with the laser launch telescope and it only detects airplanes close to the laser beam. Upon detection, it automatically stops the laser by closing a fast shutter located in the beam (safety shutter).
- Two all-sky cameras working in the visible to provide complete sky coverage. Its purpose is to detect a moving airplane and to send a warning if it is coming toward the laser beam. The cameras will be located outside the telescope in weatherproof units and their signals will be sent to the telescope for analysis.
- Radar feed from local air traffic agency. This system would provide a display with airplane location using the radar data available from different local airports.

The aircraft detection controller will be developed by Gemini Observatory and is currently under design.

Lastly, for satellite avoidance, the plan is to use the US Space Command Laser Clearinghouse Program.

## 4. CONTROL OF THE OPTO-MECHANICAL ASSEMBLIES

### 3.1 BTO/LLT Controller

The Beam Transfer Optics (BTO) is the MCAO sub-system, which brings the 5 laser beams from the Laser System to the Laser Launch Telescope (LLT) mounted behind the telescope secondary mirror. The BTO/LLT Controller is responsible for managing all of the opto-mechanical devices associated with the BTO and LLT under the direct control of the MCAO sequencer. This controller is developed by the Gemini Observatory. Preliminary design is available on the web[6].

It will control a total of (a) 27 servo motors; (b) 3 AC motors; (c) 2 stepper motors; (d) 10 piezzo actuators; (e)

BTO diagnostic wavefront sensors (2 cameras); (f) 4 closed loops (one at 800Hz rate, the other ones at slower rates as 1Hz). The Controller will be a full EPICS/VxWorks system[7].

The following architecture has been defined to satisfy the BTO/LLT control and software requirements. The Controller will be implemented using two separate CPU boards. The first CPU board will provide the EPICS interface and all device control. This board will also implement the slow closed loop algorithms. The second CPU board (non EPICS) will be dedicated to the fast closed loop at 800Hz. The CPU boards will be single PPC boards from Motorola.

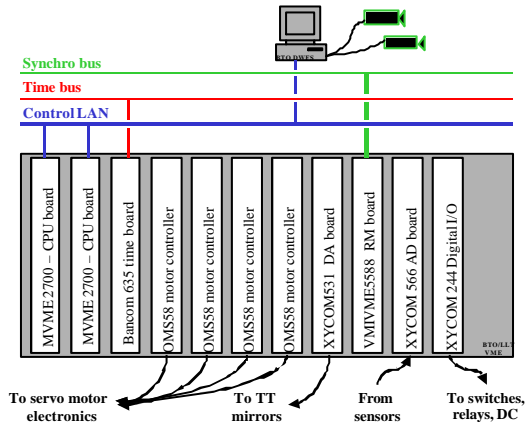

Figure 2: BTO/LLT Controller hardware architecture

### 3.2 AOM Controller

The Adaptive Optics Module (AOM) includes all of the optics, sensors and electronics needed to compensate the input f/16 science beam and relay it to a science instrument at f/33. These components include the principal elements of the real-time MCAO control loop: the 3 DMs, the TTM, 5 high-order LGS wavefront sensors, 3 tip-tilt NGS wavefront sensors. The AOM Controller is responsible for managing all of the opto-mechanical devices associated with the AOM under the direct control of the MCAO sequencer except the DMs, TTM and the readout of the NGS and LGS WFS. This controller is developed by the Gemini Observatory. Preliminary design is available on the web[6].

It will control a total of (a) 32 servo motors; (b) 9 DC motors; (c) 10 piezzo actuators; (d) 5 closed loops at slow rate (1Hz or 0.1Hz). The Controller will be a full EPICS/VxWorks system.

The following architecture has been defined to satisfy the AOM control and software requirements. The CPU board will be single PPC board from Motorola. This board will handle the control of all the devices and all the slow closed loops.

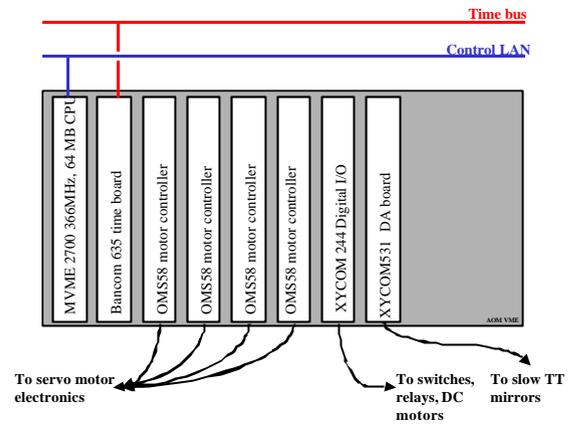

Figure3: AOM Controller hardware architecture